\documentclass{PoS}

\newcommand{\rSXPT}{rS$\chi$PT}

\title{Taste breaking effects in scalar meson correlators}

\author{C. Bernard \\  
      Department of Physics, Washington University, St.~Louis, MO 63130, USA\\
      E-mail: \email{cb@lump.wustl.edu}}

\author{\speaker{C. DeTar} and Ziwen Fu\\
   Department of Physics, University of Utah, Salt Lake City, UT 84112, USA\\
   E-mail: \email{detar@physics.utah.edu} and \email{kevin@physics.utah.edu}}

\author{S. Prelovsek\\
Department of Physics, University of Ljubljana, Jadranska 19, Ljubljana, Slovenia \\
and J.~Stefan Institute, Jamova 39, PO Box 300, 1001, Ljubljana, Slovenia\\
 E-mail: \email{Sasa.Prelovsek@ijs.si}}

\abstract{ As a consistency check of the staggered-fermion fourth-root
approximation, we analyze the $a_0$ and $f_0$ correlators, including
the effects of two-meson taste-multiplet intermediate states.  Rooted
staggered chiral perturbation theory describes the contributions from
the pseudoscalar taste multiplets in terms of only a few low energy
constants, which have all been previously determined by the MILC
collaboration.  In previous work one of us (Prelovsek) showed that the
two-meson ``bubble'' contributions could explain the observed
anomalies in the lattice data for the isovector $a_0$ channel.  In the
present work we extend this analysis to the $f_0$ channel.  On a MILC
collaboration lattice ensemble at 0.12 fm with $2+1$ flavors of
Asqtad-improved staggered fermions, we have made new measurements of
correlators in both channels for a variety of momenta.  A fit to these
correlators gives low energy constants that are reasonably consistent
with previous determinations by the MILC collaboration.}

\FullConference{XXIV International Symposium on Lattice Field Theory\\
		 July 23-28 2006\\
		 Tucson, Arizona, USA}

\ShortTitle{Scalar meson correlators}

\begin{document}

\section{Introduction}

When up and down quark masses are sufficiently light, the isovector
($a_0$) and isosinglet ($f_0$) scalar meson correlators are dominated
at large distances by two-body states composed of $\pi$, $K$, and
$\eta$.  In the lattice staggered fermion formulation, residual
fermion doubling results in a sixteen-member taste multiplet for each
meson.  When this symmetry is broken at nonzero lattice spacing, the
multiplet is split, resulting in a proliferation of two-body
intermediate states, thereby complicating the analysis of the
correlators.  To make matters worse, many of these states are lattice
artifacts with unphysical masses, and some have ghost (negative)
weights.  Thus we see nonlocalities in the form of unphysical
long-range contributions to the correlators, and we see violations
of unitarity in the form of unphysical and negative norm intermediate
states.  Such artifacts are expected to disappear in the continuum
limit.

Fortunately, these complexities are described by staggered chiral
perturbation theory (S$\chi$PT).  The rooted version of this theory
(\rSXPT) \cite{Aubin:2003mg,Bernard:2006zw} provides a strict
framework in which to analyze these two-body contributions in terms of
just a few low energy constants -- constants that are also determined
in an independent staggered chiral analysis of the pseudoscalar meson
masses and decay constants \cite{Aubin:2004fs}.

Thus the study of scalar correlators provides an explicit example of
the anticipated diseases of the fourth-root approximation and the
manner in which they are cured in the continuum limit.  Further, it
provides another test of the ability of \rSXPT\ to model rooted
staggered fermion QCD.  Previous tests include the aforementioned fits
to the pseudoscalar meson masses and decay constants and a comparison
of the topological susceptibility measured in rooted staggered fermion
QCD and in \rSXPT\cite{Billeter:2004wx}.

\section{Pseudoscalar meson multiplet}

\begin{figure}[tb]
\centerline{
\includegraphics[width=.5\textwidth]{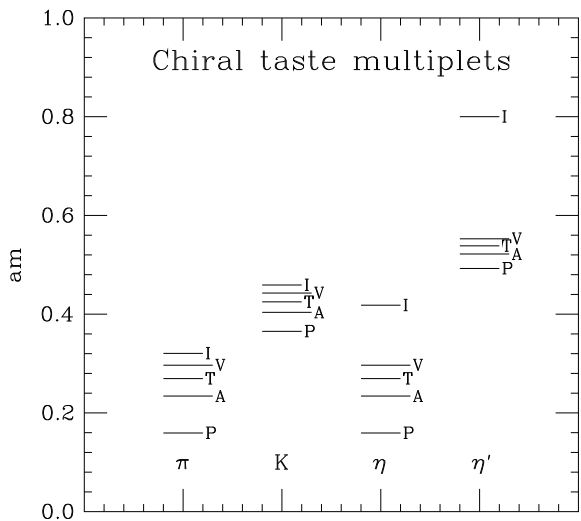}
}
\caption{Taste multiplets for $\pi$, $K$, $\eta$, and $\eta^\prime$
  for the study ensemble. The letters indicate the taste symmetry
  assignments, grouped as predicted by Lee and Sharpe
  \protect\cite{Lee:1999zx}.}
\label{fig:meson_spectrum}
\end{figure}

The $a_0$ channel was studied in recent years in staggered fermion QCD
by the MILC collaboration \cite{Aubin:2004wf} and UKQCD collaboration
\cite{Gregory:2005yr}.  Both groups found that the correlator appeared
to contain states with energies well below possible combinations of
known mesons.  At Lattice 2005 one of us showed that \rSXPT\ provides
a simple explanation \cite{Prelovsek:2005rf,Prelovsek:2005qc}.  In
Fig.~\ref{fig:meson_spectrum} we show taste splittings of the
low-lying pseudoscalar mesons for the MILC $a = 0.125$ fm ensemble, as
predicted in \rSXPT.  In the case of the pion the predictions were
confirmed in lattice measurements \cite{Aubin:2004wf,Bernard:2001av}.
Of particular concern are the $\eta$ and $\eta^\prime$ multiplets.
They are mixed by the anomaly.  But the anomaly is a taste singlet.
So only the taste-singlet members are proper candidates for the
physical states.  The other taste members are unmixed at tree level.
We have arbitrarily labeled the lighter member $\eta$ and heavier
member $\eta^\prime$.  The taste-axial-vector and vector hairpin
couplings mix the $\eta$'s weakly.  We ignore that effect in the
figure, but take it fully into account in our analysis.

Now, let us examine the various two-meson contributions to the $a_0$
correlator.  In proper QCD the lowest intermediate state has mass
$m_\pi+m_\eta$. In staggered QCD the physical $a_0$ is a taste
singlet, so we need to examine the various two-meson contributions to
the taste-singlet correlator.  Taste symmetry requires that both
pseudoscalar mesons in an intermediate state have the same taste
assignment.  In particular, the $\pi\eta$ intermediate states include
a taste-pseudoscalar pion and a taste-pseudoscalar eta with a mass
equal to twice the Goldstone pion mass.  Other taste pairs are
similarly light.  Thus we expect an anomalous low energy contribution
to the correlator.

\section{Threshold weights}

Explicit expressions were presented in
\cite{Prelovsek:2005rf,Prelovsek:2005qc} for the two-meson
contribution to the $a_0$ correlator, including all tastes.  We have
extended this analysis to the $f_0$ correlator \cite{Paper2006}. (See
also Sec.~VI of Ref.~\cite{Bernard:2006zw} for a description of the
result in the one-flavor case.)  The expressions (not written here)
consist of several two-body threshold contributions with weights
depending on the number of flavors and the replica factor
\cite{Aubin:2003rg} needed to correct for the unwanted taste
multiplicity.  If we ignore mixings induced by the taste axial-vector
and vector hairpins, the weights for the physical $2+1$ flavor case
are given in Table \ref{tab:threshold21}.  In the continuum limit in
the $a_0$ channel, all thresholds but the taste-singlet $\pi\eta$
become degenerate and the axial-vector and vector hairpin mixings
vanish.  As is evident from the table, the weights add up to zero.
The ghost threshold has eaten the unphysical thresholds, and only the
physical taste-singlet $\pi\eta$ survives.  In the $f_0$ channel, all
the $\pi\pi$ thresholds become degenerate leaving a proper weight of 3
for the three charge assignments for the physical pions.

For the single-flavor case there is no Goldstone boson, since the one
pseudoscalar is lifted by the anomaly.  Yet at nonzero lattice spacing
there is a multiplet of near Goldstone bosons composed of taste
components that are unmixed by the anomaly.  In the continuum limit
the two-body thresholds for these states become degenerate and the
weights add to zero in the $f_0$ channel as can be seen from Table
\ref{tab:threshold1}.  Thus the unphysical states decouple from this
channel.

\begin{table}[tb]
  \caption{Two-body threshold weights for $2+1$ flavors in the $a_0$
    and $f_0$ channels. The $a_0$ weights apply to the $\pi\eta$
    contributions for tastes $I,P,V,A,T$, and the $f_0$ weights apply to
    the $\pi\pi$ contributions.  Here the taste-singlet $\eta$ is the
    physical state. For the $a_0$ the weight $I0$ denotes the
    contribution from the taste-singlet pion plus the bare
    taste-singlet $\eta$ (unshifted by the anomaly).  For the $f_0$
    the weight $I0$ denotes the contribution from two bare
    taste-singlet etas. \label{tab:threshold21}}
\begin{center}
  \begin{tabular}{lrr}
    taste & $a_0$ & $f_0$ \\
     \hline
   $I$   & $ 2/3 $ & $ 1/4 $ \\
   $I0$  & $-15/8$ & $ -1 $ \\
   $V$   & $  4/8$ & $ 4/4$ \\
   $T$   & $  6/8$ & $ 6/4$ \\
   $A$   & $  4/8$ & $ 4/4$ \\
   $P$   & $  1/8$ & $ 1/4$
  \end{tabular}
\end{center}
\end{table}

\begin{table}[tb]
  \caption{Two-body threshold weights for the single flavor case $f_0$
  channel. The notation is the same as Table \protect\ref{tab:threshold21}.
   \label{tab:threshold1}}
\begin{center}
  \begin{tabular}{lr}
    taste & $f_0$ \\
     \hline
   $I0$  & $ -15/8 $ \\
   $V$   & $  4/8$   \\
   $T$   & $  6/8$   \\
   $A$   & $  4/8$   \\
   $P$   & $  1/8$  
  \end{tabular}
\end{center}
\end{table}

\section{Fits and Results}

Measurements were done on a single ensemble of 510 gauge
configurations of dimension $24^3\times 64$, generated by the MILC
collaboration with $2+1$ flavors of staggered quarks of bare mass
$am_{ud} = 0.005$ and $am_s = 0.05$ at a lattice spacing of
approximately $a = 0.125$ fm.  For this ensemble the pi/rho mass ratio
is 0.303.

We measured the correlator of scalar densities $\bar \psi \psi$ for
the isovector and isoscalar channels and for five low-lying total
momenta $p$.  The correlators were fit to the expressions
\begin{eqnarray*}
C_{a0}(p,t) &=& f_{{\rm meson},a0}(p,t)  + f_{{\rm bubble},a0}(p,t) \\
C_{f0}(p,t) &=& f_{{\rm meson},f0}(p,t)  + f_{{\rm bubble},f0}(p,t)
\end{eqnarray*}
and where the single-meson contribution is modeled by a single pole of
each parity,
\begin{eqnarray*}
  f_{{\rm meson},a0}(p,t) &=& b_{a0}(p)\exp[-E_{a0}(p) t] + 
   b_{\pi,A}(p)(-)^t \exp[-E_{\pi,A}(p)t] + (N_t - t) \\
  f_{{\rm meson},f0}(p,t) &=& c_0 \delta_(p,0) + b_{f0}(p)\exp[-E_{f0}(p)t] + 
   b_{\eta,A}(p)(-)^t \exp[-E_{\eta,A}(p)t] + (N_t - t) ,
\end{eqnarray*}
and where the ``bubble'' or two-meson contributions are dictated by
\rSXPT \cite{Prelovsek:2005qc,Prelovsek:2005rf,Paper2006}.  The meson
contributions require 13 parameters (four for the masses, two each for
the momentum dependence of the four coefficients, and one for the
zero-momentum vacuum disconnected term in the $f_0$).  For each meson
energy we assume a standard dispersion relation $E(p)^2 = m^2 + p^2$.
The meson masses in the two-meson bubble contributions are fixed from
other spectroscopic studies plus the known taste splittings
\cite{Aubin:2004wf,Aubin:2004fs}.  There remain three low energy
constants from \rSXPT, namely,
\begin{displaymath}
   \mu = m_\pi^2/(2 m_{u,d}), \ \  \delta_A = a^4 \delta_A^\prime, 
   \ \ \delta_V = a^4 \delta_V^\prime
\end{displaymath}
All three parameters were previously obtained from a study of meson
masses and decays \cite{Aubin:2004fs}.  We found it necessary to fix
(through a Bayesian prior) the vector hairpin constant $\delta_V$ to
the value obtained in that study.  The remaining two constants were
then adjusted with only loose priors.

\begin{figure}[ht]
\centerline{
\includegraphics[width=.6\textwidth]{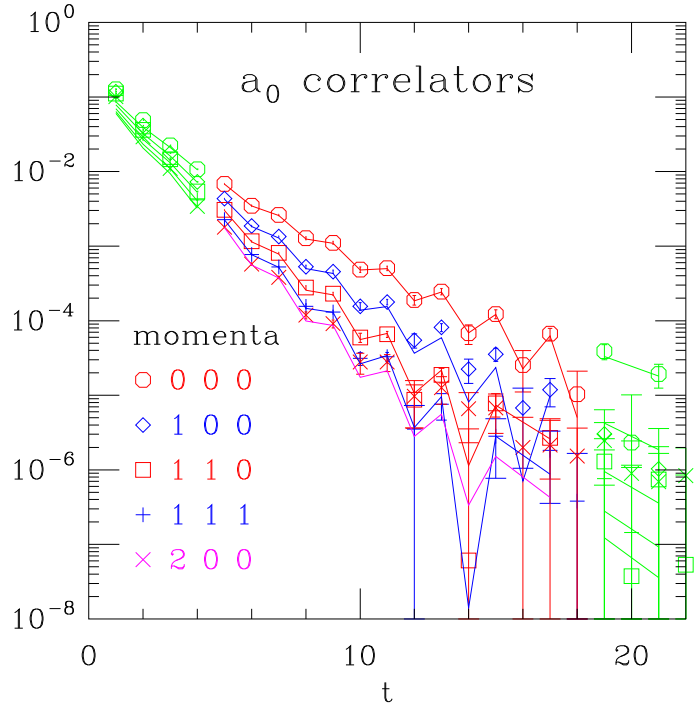}
}
\caption{Fit to the $a_0$ correlator for five momenta.
\label{fig:plot_a0_ln}}
\end{figure}

\begin{figure}
\centerline{
\includegraphics[width=.6\textwidth]{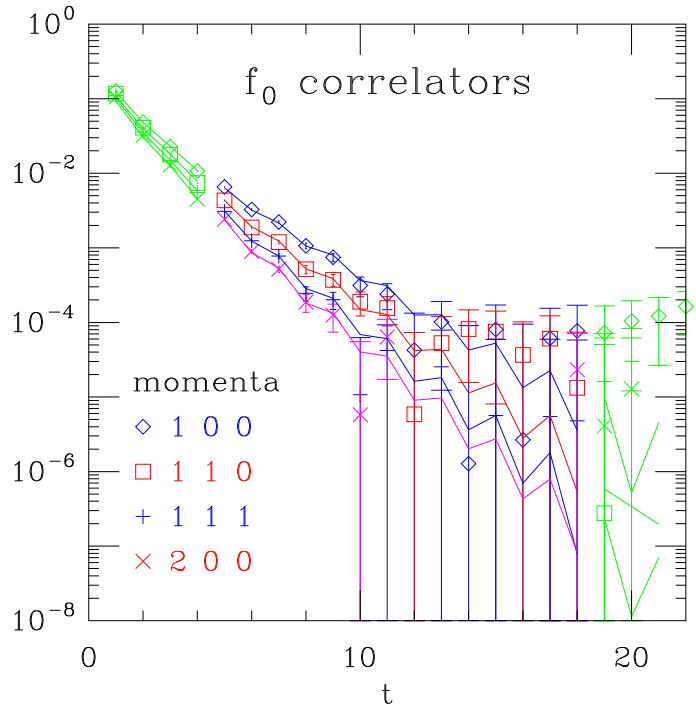}
}
\caption{Fit to the $f_0$ correlator for four momenta.
\label{fig:plot_f0_ln}}
\end{figure}

\begin{figure}
\centerline{
\includegraphics[width=.6\textwidth]{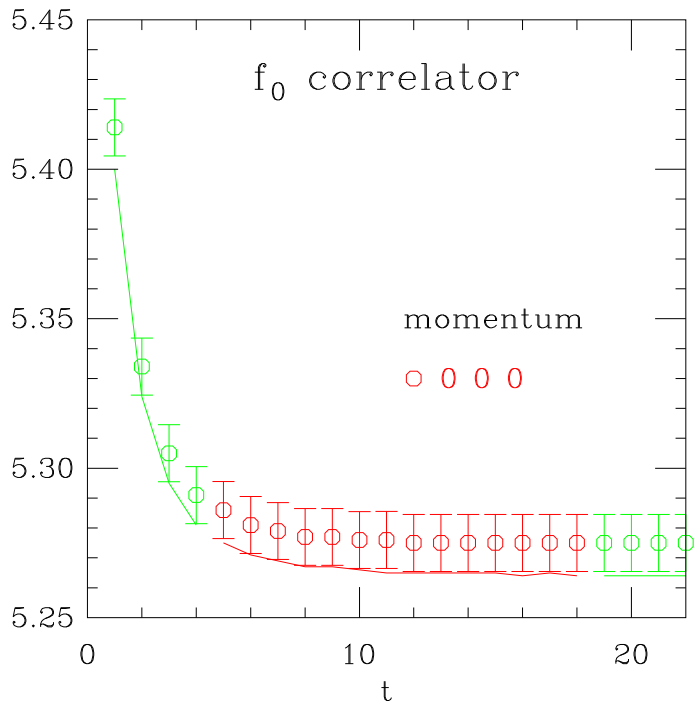}
}
\caption{Fit to the $f_0$ correlator for zero momentum.
\label{fig:plot_f0_000}}
\end{figure}

The resulting best fits are shown in Figs.~\ref{fig:plot_a0_ln},
\ref{fig:plot_f0_ln}, and \ref{fig:plot_f0_000}.

\begin{table}[htb]
  \caption{Comparison of our fit parameters for the \rSXPT\ low energy
    constants with results from \protect\cite{Aubin:2004fs} 
  \label{tab:results}}
\begin{center}
  \begin{tabular}{lrr}
    quantity  & our result & meson mass, decay \\
  \hline
    $r_1 m_\pi^2/(2 m_{u,d})$ & 8.2(1.1) & 6.7 \\
    $\delta_V              $ & (prior)  & $-0.016(23)$ \\
    $\delta_A              $ & $-0.056(7)$ & $-0.040(6)$
  \end{tabular}
\end{center}
\end{table}

The agreement is worse if we used the value of $r_1\mu$ in the
three-flavor chiral limit ($\approx 4.5$) from those fits, rather than
the higher-order $m_\pi^2/(2 m_{ud})$, suggesting, perhaps, that a
higher order calculation of the bubble contribution might improve the
agreement.

\section{Conclusion}

We have found that rooted staggered chiral perturbation theory
together with a couple of explicit low-lying mesons provides a
reasonably consistent accounting of the $a_0$ and $f_0$ correlators
for several low momenta.  The spurious long-range spectral components
are lattice artifacts, which the theory predicts will disappear in the
continuum limit.

Computing support from the Utah Center for High Performance Computing
is gratefully acknowledged.  This work was supported by the US NSF and
DOE.  We thank the MILC collaboration for the use of its lattice ensembles.

\end{document}